\documentclass[12pt]{article}%
\usepackage{amsmath,latexsym}
\usepackage{graphicx}
\usepackage{amsmath}
\usepackage{amsfonts}
\usepackage{amssymb}%
\begin{document}

\title{{Comparing $f(R)$ modified gravity and
   noncommutative geometry in the context of dark
   matter and traversable wormholes: a survey}}
   \author{
Peter K. F. Kuhfittig*\\  \footnote{kuhfitti@msoe.edu}
 \small Department of Mathematics, Milwaukee School of
Engineering,\\
\small Milwaukee, Wisconsin 53202-3109, USA}

\date{}
 \maketitle

\begin{abstract}\noindent
Noncommutative geometry, as conceptualized by
Nicolini, Smailagic, and Spallucci, may be
viewed as a slight modification of Einstein's
theory.  The same can be said for $f(R)$
modified gravity for an appropriate choice
of the function $f(R)$.  Since such an
$f(R)$ could be determined from the
noncommutative-geometry background, these
gravitational theories make very similar
predictions in the discussion of (a) dark
matter and (b) traversable wormholes; they
can therefore be taken as equally viable
models.   

\end{abstract}

\section{Introduction}\label{E:introduction}

Both $f(R)$ gravity and noncommutative geometry
may be viewed as modifications of Einstein's
theory.  For the former, the field equations
can be derived from the action
\begin{equation}\label{E:f(R)}
   S_{f(R)}=\frac{1}{2\kappa}
      \int\sqrt{-g}\,f(R)\,d^4x,
\end{equation}
where $R$ is the Ricci scalar, $f(R)$ is a
nonlinear function of $R$, and $\kappa =
8\pi G$.   If $f(R)\equiv R$, we recover the
Hilbert-Einstein action
\begin{equation}\label{E:HE}
  S_{\text{HE}}=\frac{1}{2\kappa}
     \int\sqrt{-g}\,R\,d^4x,
\end{equation}
showing that the effect of modifying
Einstein's theory can be quite small.  In
other words, a proper choice of $f(R)$
can lead to what will be referred to as
a ``slightly modified gravitational
theory."

Another example of a small effect is
noncommutative geometry, as described in
Refs. \cite{SS03, NSS06, NS10}.
Noncommutativity replaces point-like
objects by smeared objects with the aim of
eliminating the divergences that normally
appear in general relativity.  The need
for spacetime quantization was first proposed
by Snyder \cite{hS47}.  (For more recent
developments, see Refs. \cite{sF20,
FM20}.)  Further motivation is provided by
an important outcome of string theory, the
realization that coordinates may become
noncommutative operators on a $D$-brane
\cite{eW96, SW99}.  As a consequence,
spacetime can be encoded in the commutator
$[\textbf{x}^{\mu},\textbf{x}^{\nu}]
=i\theta^{\mu\nu}$, where $\theta^{\mu\nu}$
is an antisymmetric matrix that determines the
fundamental cell discretization of spacetime
in the same way that Planck's constant
discretizes phase space \cite{NSS06}.  The
physical consequences of a
noncommutative-geometry background appear to
be quite small, although the effect on the
Einstein field equations remains to be seen
(Section \ref{S:noncommutative}).

The purpose of this survey is to compare the
two gravitational theories in the context of
dark matter and traversable wormholes.  Both
theories are characterized by small effects,
ultimately resulting in very similar
predictions.

\section{Dark matter}
\subsection{$f(R)$ modified gravity}
   \label{S:modified1}
Returning now to $f(R)$ modified gravity,
renewed interest in the theory arose from
attempts to explain dark matter, as well
as the late-time accelerated expansion of
the Universe.  For further discussion,
see Refs. \cite{NO07, BHL, CDTT04, fL08, SF10}.

Since we are primarily interested in small
effects, we need to define ``slightly modified
gravitational theory," as proposed in Ref.
\cite{pK16}.  To that end, we start with
a static and spherically symmetric line
element using units in which $c=G=1$:
\begin{equation}\label{E:line1}
ds^{2}=-e^{2\Phi(r)}dt^{2}+\frac{dr^2}{1-m(r)/r}
+r^{2}(d\theta^{2}+\text{sin}^{2}\theta\,
d\phi^{2}),
\end{equation}
where $m(r)$ is the effective mass inside a
sphere of radius $r$ with $m(0)=0$ and
$\text{lim}_{r\rightarrow\infty}\frac{m(r)}
{r}=0$ \cite{MTW}.  Next, we list the
gravitational field equations in the form
used by Lobo and Oliveira \cite{LO09}
(replacing $b(r)$ by $m(r)$ for notational
convenience).  We also assume that $\Phi'(r)
\equiv 0$; otherwise, according to Ref.
\cite{LO09}, the analysis becomes intractable.
The equations are
\begin{equation}\label{E:Lobo1}
   \rho(r)=F(r)\frac{m'(r)}{r^2},
\end{equation}
\begin{equation}\label{E:Lobo2}
   p_r(r)=-F(r)\frac{m(r)}{r^3}
   +F'(r)\frac{rm'(r)-m(r)}{2r^2}
   -F''(r)\left(1-\frac{m(r)}{r}\right),
\end{equation}
and
\begin{equation}\label{E:Lobo3}
   p_t(r)=-\frac{F'(r)}{r}\left(1-\frac{m(r)}{r}
   \right)+\frac{F(r)}{2r^3}[m(r)-rm'(r)],
\end{equation}
where $F=\frac{df}{dR}$.  The Ricci
curvature scalar is given by
\begin{equation}\label{E:R}
   R(r)=\frac{2m'(r)}{r^2}.
\end{equation}
Observe that the field equations reduce to
the Einstein field equations for $\Phi'(r)
\equiv 0$ whenever $F \equiv 1$.  We can now
readily define the notion of slightly
modified gravity by assuming that $F(r)$
remains close to unity and relatively flat;
in other words, both $F'(r)$ and $F''(r)$
remain relatively small in absolute value
\cite{pK16}.

Before continuing, we need to recall that
one objective in $f(R)$ modified
gravity is to explain the flat galactic
rotation curves without assuming the
existence of dark matter.  It is known that
in the outer region of the halo, test particles
move with constant velocity in a circular path
\cite{MGN, Nandi, fR11, fR12, fR14}.
According to Ref. \cite{Nandi}, for a test
particle with four-velocity $U^{\alpha}=
dx^{\alpha}/d\tau$, we have
\begin{equation}
\left(\frac{dr}{d\tau}\right)^2=E^2+V(r),
\end{equation}
where $E$ is the relativistic energy.  An
orbit $r=a$ is stable if
\begin{equation}
   \left.\frac{d^2V}{dr^2}\right|_{r=a}<0
\end{equation}
and unstable if
\begin{equation}
   \left.\frac{d^2V}{dr^2}\right|_{r=a}>0.
\end{equation}
It is shown in Ref. \cite{pK14} that a
slightly modified gravitational theory as
defined in Ref. \cite{pK16} is enough to
yield stable galactic orbits without the
need for dark matter.  That dark matter
is a geometric effect in $f(R)$ gravity
had already been shown in Ref. \cite{BHL}.

\subsection{Noncommutative geometry}
   \label{S:noncommutative}
As noted in Sec. \ref{E:introduction},
noncommutative geometry replaces point-like
structures by smeared objects.  A direct way
to model this behavior is to use a Gaussian
distribution of minimal length $\sqrt{\beta}$
rather than the Dirac delta function
\cite{NSS06, pK08}.  An equivalent but
simpler approach is to assume that the
energy density of a static and spherically
symmetric and particle-like gravitational
source has the form \cite{NM08, pK15}
\begin{equation}\label{E:rho}
  \rho(r)=\frac{M\sqrt{\beta}}
     {\pi^2(r^2+\beta)^2},
\end{equation}
showing that the mass $M$ of a particle
is diffused throughout the region of
linear dimension $\sqrt{\beta}$ due to
the uncertainty.  An important observation
in noncommutative geometry is that
whenever we make use of Eq. (\ref{E:rho}),
we can keep the standard forms of the
Einstein field equations for the simple
reason that the Einstein tensor retains
its original form; only the stress-energy
tensor is modified \cite{NSS06}.  It
follows that the length scales can be
macroscopic.  It is also brought out in
Ref. \cite{NSS06} that noncommutative
geometry is an intrinsic property of
spacetime and does not depend on any
particular feature such as curvature.

Returning to the dark-matter hypothesis,
although first introduced in the 1930's,
the implications thereof were not fully
recognized until the 1970's when it was
observed that galaxies exhibit flat
rotation curves, i.e., constant
tangential velocities sufficiently far
from the galactic center \cite{RTF80}.
To recall why, suppose $m_1$ is the
mass of a star, $v$ its constant
velocity, and $m_2$ the mass of
everything else.  Multiplying $m_1$ by
the centripetal acceleration yields
\begin{equation}
   m_1\frac{v^2}{r}=m_1m_2\frac{G}{r^2},
\end{equation}
where $G$ is Newton's gravitational
constant.  Reverting to geometrized
units $(c=G=1)$, we obtain the linear
form
\begin{equation}
   m_2=rv^2.
\end{equation}
This form essentially characterizes the
dark-matter hypothesis.  For the purpose
of this survey, the most important
conclusion is drawn in Ref. \cite{pK17}:
the linear relationship can be attributed
to the noncommutative-geometry background.
That noncommutative geometry can account
for the flat rotation curves without the
need for dark matter is also shown in
Refs. \cite{RKCUR, KG14}.

\section{Traversable wormholes}

Wormholes are handles or tunnels that link
widely separated regions of our Universe or
different universes altogether.  Morris and
Thorne \cite{MT88} proposed the following
static and spherically symmetric line element
for a wormhole spacetime:
\begin{equation}\label{E:line}
ds^{2}=-e^{2\Phi(r)}dt^{2}+\frac{dr^2}{1-b(r)/r}
+r^{2}(d\theta^{2}+\text{sin}^{2}\theta\
d\phi^{2}).
\end{equation}
Here $b=b(r)$ is called the \emph{shape function}
and $\Phi=\Phi(r)$ is called the \emph{redshift
function}, which must be everywhere finite
to prevent the occurrence of an event horizon.
For the shape function, we must also have
$b(r_0)=r_0$, where $r=r_0$ is the radius of
the \emph{throat} of the wormhole.  An
important geometric requirement is the
\emph{flare-out condition} at the throat:
$b'(r_0)<1$, while $b(r)<r$ near the throat.
The flare-out condition can only be met by
violating the null energy condition (NEC)
\begin{equation}
  T_{\alpha\beta}k^{\alpha}k^{\beta}\ge 0
\end{equation}
for all null vectors $k^{\alpha}$, where
$T_{\alpha\beta}$ is the stress-energy
tensor.  Matter that violates the NEC is
referred to as ``exotic" in Ref. \cite{MT88}.
In particular, for the outgoing null vector
$(1,1,0,0)$, the violation has the form
\begin{equation}
   T_{\alpha\beta}k^{\alpha}k^{\beta}=
   \rho +p_r<0.
\end{equation}
Here $T^t_{\phantom{tt}t}=-\rho(r)$, where
$\rho(r)$ is the energy density,
$T^r_{\phantom{rr}r}= p_r(r)$ is the
radial pressure, and
$T^\theta_{\phantom{\theta\theta}\theta}=
T^\phi_{\phantom{\phi\phi}\phi}=p_t(r)$
is the lateral pressure.  For the wormhole
spacetime, another important physical
property is asymptotic flatness, which
demands that
$\text{lim}_{r\rightarrow\infty}\Phi(r)=0$
and $\text{lim}_{r\rightarrow\infty}b(r)/r=0$.

\subsection{$f(R)$ modified gravity}
   \label{S:modified2}
To apply $f(R)$ modified gravity to
wormholes, we need to replace $m(r)$ in
Eqs. (\ref{E:Lobo1})-(\ref{E:Lobo3}) by
$b(r)$, as in Ref. \cite{LO09}.  While
the shape function $b(r)$ has the usual
properties, the flare-out condition no
longer implies that the NEC has been
violated, as in classical general
relativity.  In fact, it is asserted
by Lobo and Oliveira in Ref. \cite{LO09}
that for the material threading the
throat of a wormhole, the NEC can be
met, thereby allowing the use of ordinary
(nonexotic) matter in the modified theory.
Imposing the conditions $\rho +p_r\ge 0$
and $\rho\ge 0$, it now follows from
Eqs. (\ref{E:Lobo1}) and (\ref{E:Lobo2})
that the function $F$ must be positive
and must satisfy the following
conditions at the throat:
\begin{equation}
   \frac{Fb'}{r^2}\ge0
\end{equation}
and
\begin{equation}
   \frac{(2F+rF')(b'r-b)}{2r^2}-F''
   \left(1-\frac{b}{r}\right)\ge0.
\end{equation}
To complete the proof of the above
assertion, we also need to show that
the NEC is met for all null vectors
\begin{equation}
   (1,a,b,c), \,\,\text{where}\,\,
   0\le a,b,c\le 1\,\,\text{and}\,\,
      a^2+b^2+c^2=1.
\end{equation}
First, from Eq. (\ref{E:Lobo3}) we have,
at the throat, $\rho +p_t\ge 0$, $F$ being
positive.  Since $\rho=(a^2+b^2+c^2)\rho$,
we can write $\rho +a^2p_r+b^2p_t+c^2p_t$
in the form
\begin{multline*}
   a^2(1,0,0,0)+b^2(1,0,0,0)+c^2(1,0,0,0)
   +a^2(0,1,0,0)+b^2(0,0,1,0)+c^2(0,0,0,1)\\
   =a^2(1,1,0,0)+b^2(1,0,1,0)+c^2(1,0,0,1)\\
   =a^2(\rho+p_r)+b^2(\rho+p_t)+c^2(\rho+p_t)\ge 0,
\end{multline*}
showing that the NEC is satisfied for all
null vectors.

Addressing an apparent conflict with the
classical theory, it is shown in Ref. \cite{HLMS}
that to sustain a worhole, a violation of the
NEC is indeed unavoidable by referring back to
the Raychaudhury equation. According to Ref.
\cite{LO09}, such a violation can be attributed
to the higher-order curvature terms, interpreted
as a gravitational fluid.

Having established that $f(R)$ modified
gravity can support traversable wormholes,
we need to recall that we are primarily
interested in slightly modified gravitational
theories, as defined in Sec. \ref{S:modified1}.
This issue is addressed in Ref. \cite{pK16}
by showing that even a slight modification is
sufficient for sustaining a wormhole, while
avoiding a violation of the NEC for matter
threading the throat.  The analysis assumes
zero tidal forces.

\subsection{Noncommutative geometry}

In this section we return to Eq. (\ref{E:rho})
and line element (\ref{E:line}) describing a
wormhole spacetime.  Here we need to list the
Einstein field equations in classical general
relativity:
\begin{equation}\label{E:Einstein1}
  \rho(r)=\frac{b'}{8\pi r^2},
\end{equation}
\begin{equation}\label{E:Einstein2}
   p_r(r)=\frac{1}{8\pi}\left[-\frac{b}{r^3}+
   2\left(1-\frac{b}{r}\right)\frac{\Phi'}{r}
   \right],
\end{equation}
\begin{equation}\label{E:Einstein3}
   p_t(r)=\frac{1}{8\pi}\left(1-\frac{b}{r}\right)
   \left[\Phi''-\frac{b'r-b}{2r(r-b)}\Phi'
   +(\Phi')^2+\frac{\Phi'}{r}-
   \frac{b'r-b}{2r^2(r-b)}\right].
\end{equation}
Eqs. (\ref{E:rho}) and (\ref{E:Einstein1})
yield the shape function \cite{pK15}
\begin{equation}
   b(r)=\frac{4M\sqrt{\beta}}{\pi}
  \left(\frac{1}{\sqrt{\beta}}\text{tan}^{-1}
  \frac{r}{\sqrt{\beta}}-\frac{r}{r^2+\beta}-
  \frac{1}{\sqrt{\beta}}\text{tan}^{-1}
  \frac{r_0}{\sqrt{\beta}}+\frac{r_0}{r_0^2
  +\beta}\right)+r_0,
\end{equation}
where $M$ is now the mass of the wormhole.
To check the flare-out condition, we first
observe that
\begin{equation}
   b'(r)=\frac{4M\sqrt{\beta}}{\pi}
   \frac{2r^2}{(r^2+\beta)^2}>0,
\end{equation}
as required by Eq. (\ref{E:Einstein1}), but
it also follows that $b'(r)<1$ as long as
$\sqrt{\beta}\ll M$.  So the flare-out
condition is met and the NEC is automatically
violated, as we saw earlier.  As another
check,
\begin{equation}
   \rho+p_r=
  \frac{M\sqrt{\beta}}{\pi^2(r^2+\beta)^2}
  -\frac{1}{8\pi}\frac{b(r)}{r^3}<0
\end{equation}
at or near the throat since $\sqrt{\beta}
\ll 1$.  Finally, $\text{lim}_{r\rightarrow
\infty}b(r)/r=0$, showing that the wormhole
spacetime is asymptotically flat, provided that
$\text{lim}_{r\rightarrow\infty}\Phi(r)=0$.

Higher-dimensional wormholes, given a
noncommutative-geometry background, are
discussed in Ref. \cite{RKRI}.  Charged
wormholes with low tidal forces that are
also inspired by noncommutative geometry
are analyzed in Ref. \cite{KG17}.  Wormholes
that take into account both gravitational
theories, i.e., noncommutative wormholes
in $f(R)$ gravity, are discussed in Ref.
\cite{pK18a}.  The effects of the two
gravitational theories are independent
without being mutually exclusive.  As a
result, the combined effects may differ
from the individual effects.

The possible avoidance of exotic matter
at the throat is discussed in the next
section.

\section{Connecting noncommutative geometry
  to $f(R)$ modified gravity}
Both $f(R)$ modified gravity and
noncommutative geometry can help sustain
traversable wormholes and both can account
for the flat galactic rotation curves and
hence for dark matter, even if the effect
of $f(R)$ gravity is small.  The effect of
noncommutative geometry is small to
begin with, thereby suggesting a
connection between the two: according to
Ref. \cite{pK18b}, a noncommutative-geometry
background can determine the corresponding
function $f(R)$ for modeling dark matter
and may even provide a motivation for the
choice of $f(R)$.  So it follows from
Sec. \ref{S:modified2} that a
noncommutative-geometry wormhole can
in principle be constructed without
exotic matter.  This conclusion is made
explicit in Ref. \cite{pK20} by determining
the form of $f(R)$ for the wormhole
spacetime:
\begin{equation}\label{E:f(R)}
    f(R)=
    \frac{2M\sqrt{\beta}}{\pi^2}
    \frac{(\beta R+2b')\text{ln}\,
    (\beta R+2b')-\beta R}
    {\beta^2(\beta R+2b')}+C.
\end{equation}
The zero-tidal force assumption is retained
for the technical reason given in Sec.
\ref{S:modified1} but it does not constitute
a necessary condition.

\section{Summary}
This survey has pointed out that noncommutative
geometry and $f(R)$ modified gravity are
equally viable models for discussing dark
matter and traversable wormholes, in part
because the function $f(R)$ can be determined
from the noncommutative-geometry background.

Even a slightly modified $f(R)$ gravitational
theory can account for the galactic rotation
curves and hence for dark matter.  A
noncommutative-geometry
background with its characteristic small
effect can account for the same phenomenon.
So dark matter can be viewed as a geometric
effect in modified gravity.  It is also
noted that both gravitational theories are 
not only able to support traversable 
wormholes, they allow the throat to be 
threaded with nonexotic matter.  The 
unavoidable violation of the NEC can be 
attributed to the higher-order curvature 
terms in the modified theory.


\begin{thebibliography}{20}
\bibitem{SS03}A. Smailagic and E. Spallucci, ``Feynman
   path integral on a non-commutative plane," J. Phys. A
   \textbf{36}, L-467-L-471 (2003).
   \emph{}
\bibitem{NSS06}P. Nicolini, A. Smailagic, and E. Spallucci,
   ``Noncommutative geometry inspired Schwarzschild
   black hole," Phys. Lett. B \textbf{632}, 547-551 (2006).
\bibitem{NS10}P. Nicolini and E. Spallucci,
   ``Noncommutative geometry-inspired dirty black holes,"
   Class. Quant. Grav. \textbf{27}, 015010 (2010).
\bibitem{hS47}H. S. Snyder, ``Quantized space-time,"
   Phys. Rev. \textbf{71}, 38-41 (1947).
\bibitem{sF20}S. A. Franchino-Vinas, ``Asymptotic
   freedom for $\lambda\Phi^4_{\star}$ QFT in
   Snyder-de Sitter space," Eur. Phys. J. C
   \textbf{80}, 382 (2020).
\bibitem{FM20}S. A. Franchino-Vinas and S. Mignemi,
   ``Casimir effect in Snyder space," Nucl. Phys. B
   \textbf{959}, 115152 (2020).
\bibitem{eW96}E. Witten, ``Bound states of strings
   and $p$-branes," Nucl. Phys. B \textbf{460},
   335-350 (1996).
\bibitem{SW99}N. Seiberg and E. Witten, ``String theory
   and noncommutative geometry," J. High Energy Phys.
   \textbf{1999}, 032 (1999).
\bibitem{NO07}S. Nojiri and S. D. Odintsov, ``Introduction
   to modified gravity and gravitational alternative for
   dark matter," Int. J. Geom. Meth. Mod. Phys.
   \textbf{4}, 115-145 (2007).
\bibitem{BHL}C. B\"{o}hmer, T. Harko, and F. S. N. Lobo,
   ``Dark matter as a geometric effect in $f(R)$ gravity,"
   Astropart Phys, \textbf{29}, 386-392 (2008).
\bibitem{CDTT04}S. M. Carroll, V. Duvvuri, M. Trodden
   and M. S. Turner, ``Is cosmic speed-up due to new
   gravitational physics?" Phys. Rev. D \textbf{70},
   043528 (2004).
\bibitem{fL08}F. S. N. Lobo, ``The dark side of gravity:
   Modified theories of gravity," arXiv: 0807.1640 [gr-qc].
\bibitem{SF10}T. P. Sotiriou and V. Faraoni, ``$f(R)$
   theories of gravity," Rev. Mod. Phys. \textbf{82},
   451-497 (2010).
\bibitem{pK16}P. K. F. Kuhfittig, ``A note on wormholes
in slightly modified gravitational theories," Adv. Studies
   Theor Phys. \textbf{7}, 1087-1093 (2013)
\bibitem{MTW}C. W. Misner, K. S. Thorne, and J. A. Wheeler,
   Gravitation (New York: W. Freeman and Company, 1973),
   page 608.
\bibitem{LO09}F. S. N. Lobo and M. A. Oliveira,
   ``Wormhole geometries in $f(R)$ modified theories of
   gravity," Phys. Rev. D \textbf{80}, 104012 (2009).
\bibitem{MGN}T. Matos, F. S. Guzman, and D. Nunez,
   ``Spherical scalar field halo in galaxies," Phys.
   Rev. D \textbf{62}, 0611301 (2006).
\bibitem{Nandi}K. K. Nandi, A. I. Filippov, F. Rahaman,
   S. Ray, A. A. Usmani, M. Kalam, and A. DeBenedictis,
   ``Features of galactic halo in a brane world model
   and observational constraints,"  Mon. Not. Roy.
   Astron. Soc. \textbf{399}, 2079-2087 (2009).
\bibitem{fR11}F. Rahaman, P. K. F. Kuhfittig, K.
   Chakraborty, M. Kalam, and D. Hossain, ``Modeling
      galactic halos with predominantly quintessential
      matter," Int. J. Theor. Phys, \textbf{50},
      2655-2665 (2011).
\bibitem{fR12}F. Rahaman, P. K. F. Kuhfittig, R. Amin,
   G. Mandel, S. Ray, and N. Islam, ``Quark matter as
   dark matter in modeling galactic halo," Phys. Lett. B
   \textbf{714}, 131-135 (2012).
\bibitem{fR14}R. Rahaman, R. Biswas, H. I. Fatima, and
   N. Islam, ``A new proposal for galactic dark matter:
   effect of $f(T)$ gravity," Int. J. Theor. Phys.
   \textbf{53}, 370-379 (2014).
\bibitem{pK14}P. K. F. Kuhfittig, ``A simple argument
   for dark matter as an effect of slightly modified
   gravity," Adv. Studies Theor. Phys. \textbf{8},
   349-355 (2014).
\bibitem{pK08}P. K. F. Kuhfittig, ``Macroscopic wormholes
   in noncommutative geometry," Int. J. Pure Appl. Math.
   \textbf{89}, 401-408 (2013).
\bibitem{NM08}N. Nozari and S. H. Mehdipour, ``Hawking
   radiation as quantum tunneling from a noncommutative
   Schwarzschild black hole," Class. Quant. Grav.
   \textbf{25}, 175015 (2008).
\bibitem{pK15}P. K. F. Kuhfittig, ``Macroscopic traversable
   wormholes with zero tidal forces inspired by
   noncommutative geometry," Int. J. Mod. Phys. D
   \textbf{24}, 1550023 (2015).
\bibitem{RTF80}V. Rubin, N. Thonnard, and W. K. Ford,
   ``Rotational properties of 21 SC galaxies with a
   large range of luminosities and radii, from NGC 4605/R
   = 4 kpc/ to UGC 2885/R = 122 kpc/" Astroph. J.
   \textbf{238}, 471-487 (1980).
\bibitem{pK17}P. K. F. Kuhfittig, ``Accounting for some
   aspects of dark matter and dark energy via
   noncommutative geometry," J. Mod. Phys. \textbf{8},
   323-329 (2017).
\bibitem{RKCUR}F. Rahaman, P. K. F. Kuhfittig, K.
   Chakraborty, A. A. Usmani, and S. Ray, ``Galactic
   rotation curves inspired by a noncommutative-geometry
   background," Gen. Rel. Grav. \textbf{44}, 905-916
   (2012).
\bibitem{KG14}P. K. F. Kuhfittig and V. D. Gladney,
   ``Revisiting galactic rotation curves given a
   noncommutative-geometry background," J. Mod. Phys.
   \textbf{5}, 1931-1937 (2014).
\bibitem{MT88}M. S. Morris and K. S. Thorne, ``Wormholes
   in spacetime and their use for interstellar travel: A
   tool for teaching general relativity," Amer. J. Phys.
   \textbf{56}, 395-412 (1988).
\bibitem{HLMS}T. Harko, F. S. N. Lobo, M. K. Mak, and S. V.
   Sushkov, ``Modified-gravity wormholes without exotic
   matter," Phys. Rev. D \textbf{87}, 067504 (2013).
 \bibitem{RKRI}F. Rahaman, S. Islam, P. K. F. Kuhfittig,
    and S. Ray, ``Searching for higher-dimensional
    wormholes with noncommutative geometry," Phys.
     Rev. D \textbf{86}, 106010 (2012).
\bibitem{KG17}P. K. F. Kuhfittig and V. D. Gladney,
   ``Noncommutative-geometry inspired charged wormholes
     with low tidal forces," J. Appl. Math. Phys. (JAMP)
     \textbf{5}, 574-581 (2017).
\bibitem{pK18a}P. K. F. Kuhfittig, ``Wormholes in $f(R)$
   gravity with a noncommutative-geometry background,"
   Indian J. Phys. \textbf{92}, 1207-1212 (2018).
\bibitem{pK18b}P. K. F. Kuhfittig, ``Connecting
   noncommutative geometry to $f(R)$ modified gravity,"
   New Horizons in Math. Phys. (NHMP) \textbf{2},
   62-66 (2018).
\bibitem{pK20}P. K. F. Kuhfittig,
   ``Noncommutative-geometry wormholes without exotic
   matter," Adv. Studies Theor. Phys. \textbf{14},
   219-225 (2020).





 \end{thebibliography}
\end{document}